\documentclass[12pt]{article}
\usepackage{a4wide}
\usepackage{amssymb}
\begin{document}
{\renewcommand{\thefootnote}{\fnsymbol{footnote}}
\hfill  IGPG--06/6--3\\
\medskip
\hfill hep--th/0606232\\
\medskip
\begin{center}
{\LARGE  Quantum Gravity and Higher Curvature Actions}\\
\vspace{1.5em}
Martin Bojowald$^1$\footnote{e-mail address: {\tt bojowald@gravity.psu.edu}}
and Aureliano Skirzewski$^2$\footnote{e-mail address: {\tt skirz@aei.mpg.de}}
\\
\vspace{0.5em}
$^1$Institute for Gravitational Physics and Geometry,\\
The Pennsylvania State
University,\\
104 Davey Lab, University Park, PA 16802, USA\\
\vspace{0.5em}
$^2$Max-Planck-Institut f\"ur Gravitationsphysik,
Albert-Einstein-Institut,\\
Am M\"uhlenberg 1, D-14476 Potsdam, Germany\\
\vspace{1.5em}
\end{center}
}
                                                                                
\setcounter{footnote}{0}

\newcommand{\md}{\mathrm d}

\begin{abstract}
 Effective equations are often useful to extract physical information
 from quantum theories without having to face all technical and
 conceptual difficulties. One can then describe aspects of the quantum
 system by equations of classical type, which correct the classical
 equations by modified coefficients and higher derivative terms. In
 gravity, for instance, one expects terms with higher powers of
 curvature. Such higher derivative formulations are discussed here
 with an emphasis on the role of degrees of freedom and on differences
 between Lagrangian and Hamiltonian treatments. A general scheme is
 then provided which allows one to compute effective equations
 perturbatively in a Hamiltonian formalism. Here, one can expand
 effective equations around any quantum state and not just a
 perturbative vacuum. This is particularly useful in situations of
 quantum gravity or cosmology where perturbations only around vacuum
 states would be too restrictive. The discussion also demonstrates the
 number of free parameters expected in effective equations, used to
 determine the physical situation being approximated, as well as the
 role of classical symmetries such as Lorentz transformation
 properties in effective equations. An appendix collects information
 on effective correction terms expected from loop quantum gravity and
 string theory.
\end{abstract}

\section{Introduction}

Quantum theories are usually full of considerable technical and
conceptual difficulties. Even in mechanical systems which classically
have only a finite number of degrees of freedom one has to deal with
partial differential equations for a wave function. Such equations are
more complicated to solve than the ordinary differential equations of
a classical mechanical system, and for their solutions one faces the
usual interpretational issues. In quantum field theory the situation
is correspondingly more complex since even classical field theories
have infinitely many degrees of freedom. One is then dealing with some
kind of functional differential equations in quantum field
theory. This certainly applies also to gravity which is a field
theory, unless one uses cosmological minisuperspace models with only a
finite number of degrees of freedom. In the case of gravity,
interpretational issues of the wave function related to gravity, such
as the definition of observables or the problem of time, are even more
severe.

It is therefore of considerable interest to find techniques which in
certain regimes allow one to approximate the quantum system by a
system of classical type, or {\em effective system}, by which we mean
a system whose dynamical laws are of the same mathematical class as
those of a corresponding classical system. For instance, an effective
system of quantum mechanics would describe some regime of a quantum
mechanical system by ordinary differential equations. This can be
thought of as describing the motion of a wave packet by ordinary
differential equations for the peak position rather than a partial
differential equation for the whole wave function. In semiclassical
regimes, effective equations should then be close to the classical
ones, up to corrections of order $\hbar$. More generally, however,
there can be other regimes where the classical equations are not valid
at all, but some other equations of classical type still suffice. This
is, for instance, the case if a wave packet spreads and deforms and
these deformations back-react on the motion of the peak. Quantum
properties are then important, but there could well be a finite number
of parameters describing the wave packet well enough to justify the
use of ordinary differential equations. An effective description is
thus much more general than a semiclassical one.

A method of widespread use in particular in quantum field theory is
that of low energy effective actions \cite{LowEnEffAc}, which allows
one to describe the quantum system by an effective action amending the
classical action by quantum corrections but leaving the classical
structure mainly untouched. In many cases, in particular when one
expands perturbatively around a free field theory or a harmonic
oscillator, such effective actions can be computed explicitly. The
usual definition through a Legendre transform of the generating
functional of irreducible $n$-point functions, however, looks very
non-intuitive and seems tied to the situation of perturbative quantum
field theory. Superficially, it looks unrelated to the above picture
of traveling wave packets, and it is of deceptive uniqueness: there
are usually no free parameters in the low energy effective action
while an effective system for a quantum system should depend on the
initial form of the wave packet, such as its initial spread, whose
evolution is to be described effectively.

In this article we will discuss the relation between both types of
effective systems from the point of view of a generalization of the
usual effective action picture. This more general scheme has several
advantages: (i) it does not restrict one to perturbative treatments
and in fact allows all necessary freedom one may want to include in
suitable initial states to perturb around, (ii) it applies to
Hamiltonian methods as well and thus allows a comparison between
effective pictures derived from canonical and covariant approaches to
quantum field theory and (iii) by displaying all possible free
parameters it shows the regimes where a particular effective action,
including the standard low energy one, should or should not be
applied. The last point has in particular implications for quantum
gravity and cosmology, as we will discuss in the end. (Effective
results from different quantum theories of gravity are collected in
the Appendices.) We will also see that all crucial aspects of an
effective action are already present for systems of finitely many
degrees of freedom such as cosmological models in the context of
gravity. It is thus sufficient for most purposes to use
simplifications realized in such models, while field theoretical
degrees of freedom increase the complexity considerably but do not add
many distinctive properties.

\section{Higher derivatives}

Effective actions usually contain terms of the classical action
with quantum corrections in the coefficients such as mass
renormalization or corrections in an effective potential. An example
is the effective action \cite{EffAcQM}
\begin{equation} \label{EffAc}
\Gamma_{{\rm eff}}[q]=\int
\md t\left[\left(m+\frac{\hbar U'''(q)^2}
{32m^2\left(\omega^2+\frac{U''(q)}{m}\right)^{\frac{5}{2}}}\right)
\frac{\dot q^2}{2}-\frac{1}{2}m\omega^2q^2
-U(q)-\frac{\hbar\omega}{2} \left(1+\frac{U''(q)}{m\omega^2}
\right)^{\frac{1}{2}}\right]
\end{equation}
for an an-harmonic oscillator with classical potential
$\frac{1}{2}m\omega^2q^2+U(q)$. In addition, there will also be higher
time derivative terms of the configuration variables in an expansion
of slowly varying variables.  This is often taken as an indication
that there are additional degrees of freedom, albeit still finitely
many ones, since more initial values have to be specified including
higher derivatives in time of the classical variables. This appears to
be along the lines sketched in the Introduction since a quantum system
does have more degrees of freedom than a classical one. One needs an
infinite number of parameters to specify a quantum state compared to
finitely many ones in classical mechanics. Thus, in an approximation
one has to include additional parameters to bridge the gap, bringing
one in a possible full series summation of all higher derivative terms
to an infinite number.

\subsection{Perturbation}

While this is true for any given higher derivative action, one has to
be more careful in a perturbative scheme in which the higher
derivative terms arise here. The highest order term in equations of
motion following from such an effective action will always be
multiplied by a power of the perturbation parameter such as
$\hbar$. The unperturbed classical system is thus a singular point of
the equations from a mathematical point of view
\cite{SingHighOrder}. Thus, most solutions will diverge when the
perturbation parameter goes to zero, which should clearly be avoided
for any solution one would trust within the perturbative
scheme. Keeping solutions which are not analytic at zero perturbation
parameter would violate the approximation in which effective equations
have been derived.

As an example,\footnote{See also \cite{CorrectScalar} for further
discussion of this example.} let us consider the field theory action
\begin{equation} \label{scalarL}
 S[\psi]=-\frac{1}{2}\int(\psi(\Box+\epsilon\Box^2)\psi+m^2\psi^2)\md^3x\md t
\end{equation}
of a scalar field $\psi$ which has some higher derivative correction
in a perturbation parameter $\epsilon$. The field equation is
\begin{equation} \label{eomL}
 -(\Box+\epsilon\Box^2)\psi=m^2\psi
\end{equation}
with highest order term multiplied by $\epsilon$. The unperturbed
theory ($\epsilon=0$) is thus of lower order than the perturbed one
and has less independent solutions.

In any such situation, there is a mismatch between unperturbed and
perturbed solutions, and most solutions to the perturbed equations
must be non-analytical in the perturbation parameter. For such
solutions, higher order terms contribute uncontrolled corrections such
that they have to be discarded in a perturbative treatment.  Keeping
them is inconsistent within the perturbative analysis. Only analytical
solutions are to be retained, which do exist in the right number being
the same as the number of unperturbed solutions \cite{Simon}. In our
example, for instance, we can look for plane wave solutions
$\psi(x,t)=\exp(i(Et-kx))$ which have to fulfill the dispersion
relation $E^2=k^2-\frac{1}{2}\epsilon^{-1}\pm
\frac{1}{2}\epsilon^{-1}\sqrt{1+4\epsilon m^2}$.  For
$\epsilon\ll m^{-2}$, we can expand $E^2=k^2-\frac{1}{2}\epsilon^{-1}
(1\mp(1+2\epsilon m^2-2\epsilon^2m^4))+O(\epsilon^2)$, clearly showing
the existence of two analytical solutions in $\epsilon$, for which the
unperturbed perturbation relation is just corrected by terms of the
order $\epsilon$, $E^2=k^2+m^2-\epsilon m^4+O(\epsilon^2)$, and two
non-analytical ones which have to be discarded.

There are thus no additional degrees of freedom in the sense of
solutions, although corrections in solutions also come from higher
derivative terms even in solutions analytical in the perturbation
parameter. Thus, higher derivative effective actions, when treated
consistently in a perturbative scheme, provide additional ``quantum''
degrees of freedom only implicitly. The discussion also implies that a
description by an effective action is usually more complicated than it
appears because one not only has to solve the equations of motion but
also pick the correct ones which are analytical in the perturbation
parameter \cite{SingHighOrder}. This can be particularly difficult to
do in numerical studies. Taking general solutions at face value, on
the other hand, is misleading since this contains redundant,
non-physical degrees of freedom.

\subsection{Hamiltonian picture}

The Hamiltonian picture of effective actions looks quite different
from the Lagrangian viewpoint, although as usually both pictures are
in the end equivalent.  Degrees of freedom in a Hamiltonian
formulation are given by coordinates and momenta, which are fixed from
the outset independently of the Hamiltonian as a function of phase space
variables but not their time derivatives. This is different from the
Lagrangian which is a functional of the configuration variables as
functions of time and thus determines what the independent degrees of
freedom are. If one were to perturb a Hamiltonian theory in its given
variables one would thus obtain correction terms, but no additional
degrees of freedom, not even implicitly.

In fact, the process of expanding perturbatively does not commute with
the Legendre transformation if higher time derivatives arise
\cite{CorrectScalar}. If we just expand a Hamiltonian in its classical
variables we obtain a theory different from that obtained by expanding
the corresponding action in a higher derivative expansion and then
performing the Legendre transformation to a Hamiltonian picture. This
is because a higher derivative action superficially introduces new
degrees of freedom corresponding to higher derivatives and their
momenta, which will then also occur in the Hamiltonian
picture. Starting already at the Hamiltonian level, on the other hand,
there is no obvious way to obtain new degrees of freedom.

To continue with our example, we have momenta
 $\pi_{\psi}=
\dot{\psi}-2\epsilon\Delta\dot{\psi}+\epsilon\md^3{\psi}/\md
t^3$ and $\pi_{\dot{\psi}}=-\epsilon\ddot{\psi}$ conjugate to the
independent configuration variables $\psi$ and $\dot{\psi}$. Also
here, $\epsilon$ appears as a coefficient such that, when we invert
the relations to replace $\ddot{\psi}$ and $\md^3{\psi}/\md t^3$ in
the Hamiltonian, we divide by $\epsilon$ and the Hamiltonian
\begin{equation}\label{scalarH}
 H=\int\md^3x\left(-{\textstyle\frac{1}{2}}\dot{\psi}^2+\dot{\psi}\pi_{\psi}-
{\textstyle\frac{1}{2}}\psi\Delta\psi+ {\textstyle\frac{1}{2}}m^2\psi^2+
\epsilon(\dot{\psi}\Delta\dot{\psi}+ \psi\Delta^2\psi/2)-
{\textstyle\frac{1}{2}}\epsilon^{-1}\pi_{\dot{\psi}}^2\right)
\end{equation}
of the higher derivative action is not analytic in the perturbation
parameter. Not all these terms could have been obtained from a
perturbative treatment at the Hamiltonian level. In such a case, we
could have derived at most the perturbative terms in (\ref{scalarH})
which are analytic in $\epsilon$. The analytic part of the Hamiltonian
implies the standard momenta $\pi_{\psi}=\dot{\psi}+O(\epsilon)$ to
leading order: without additional degrees of freedom, $\dot{\psi}$ and
$\pi_{\psi}$ are not independent,
$\delta\dot{\psi}/\delta\pi_{\psi}\not=0$, such that the Hamiltonian
equation of motion from the part of (\ref{scalarH}) analytic in
$\epsilon$,
\[
 \dot{\psi}=\delta H/\delta\pi_{\psi}=
-\dot{\psi}\partial\dot{\psi}/\partial\pi_{\psi}+\dot{\psi}+
\pi_{\psi}\partial\dot{\psi}/\pi_{\psi}+2\epsilon\Delta\dot{\psi}
\partial\dot{\psi}/\partial\pi_{\psi}+ O(\epsilon^2)\,,
\]
requires $\pi_{\psi}=\dot{\psi}-2\epsilon\Delta\dot{\psi}$. This
results in the perturbative Hamiltonian
\[
  H={\textstyle\frac{1}{2}}\left(\pi_{\psi}^2-\psi\Delta\psi+ m^2\psi^2+
\epsilon(2\pi_{\psi}\Delta\pi_{\psi}+ \psi\Delta^2\psi)\right)\,.
\]
In second
order form, we obtain the equation of motion
\[
 \ddot{\psi}=\Delta\psi-m^2\psi+\epsilon(\Delta^2\psi-2m^2\Delta\psi)
+O(\epsilon^2)
\]
which is different from the Lagrangian one (\ref{eomL}).

\section{Quantum degrees of freedom}

The example illustrates that the Hamiltonian picture requires more
refined methods to derive corrections mediated through higher
derivatives or some other kind of new degrees of freedom not present
in the classical system. As described in the introduction, this can be
possible by describing a trajectory of semiclassical wave
packets by effective equations. For a mechanical system of a single
degree of freedom, we have classical variables $q=\langle\hat{q}\rangle$,
$p=\langle\hat{p}\rangle$ associated with expectation values of
quantum operators. This allows us to make an identification between
some degrees of freedom of the quantum theory and the classical ones.
A wave packet, however, has more information than just expectation
values of basic operators because we have, e.g.,
$\langle\hat{q}^n\rangle\not=\langle\hat{q}\rangle^n$. Unlike the
classical situation where $q^n$ would directly be obtained by taking a
power of $q$, the expectation value of $\hat{q}^n$ has additional
information not contained in $\langle\hat{q}\rangle^n$.
This additional information can be captured in
quantum variables \cite{EffAc}
\begin{equation}
 G^{a,n}:=\langle(\hat{q}-\langle\hat{q}\rangle)^{n-a}
 (\hat{p}-\langle\hat{p}\rangle)^a\rangle_{\rm Weyl}
 \quad,\quad a=0,1,\ldots, n
\end{equation}
which are independent of the classical ones (the subscript ``Weyl''
denoting symmetric ordering of the operators). In fact, one can define
a symplectic structure, using the imaginary part of the inner product,
for these variables such that quantum mechanics is formulated on an
$\infty$-dimensional phase space. If we label states
$|\psi\rangle=\sum_jc_j|\psi_j\rangle$ by expansion coefficients $c_j$
in some orthonormal basis $|\psi_j\rangle$, such that the inner
product between two states $|\psi\rangle$ and
$|\phi\rangle=\sum_jd_j|\psi_j\rangle$ is
$\langle\psi,\phi\rangle=\sum_j\bar{c}_jd_j$, we can use real and
imaginary parts of the $c_j$ as real coordinates on the Hilbert space
and define the symplectic structure \cite{GeomQuantMech}
\[
 \Omega(\delta c_j,\delta d_k):=2\hbar{\rm Im}\langle\psi,\phi\rangle=
 2\hbar\sum_j({\rm Re}\delta c_j{\rm Im}\delta d_j- {\rm Im}\delta
 c_j{\rm Re}\delta d_j)
\]
evaluated in two vectors with components $\delta c_j$ and $\delta d_k$.
{}From this, we read off the Poisson brackets $\{{\rm Re} c_j,{\rm Im} c_k\}=
\frac{1}{2\hbar}\delta_{jk}$, and vanishing brackets between real and
imaginary parts, respectively. While the Poisson relations are most
easily determined for expansion coefficients $c_j$, these variables
are not very suitable to be split into classical and quantum parts. We
will therefore use  below the variables $q$, $p$ and $G^{a,n}$ and also
state their Poisson relations there.

In this picture, the Schr\"odinger equation for $|\psi\rangle$ is
equivalent to Hamiltonian equations of motion with the above Poisson
brackets and Hamiltonian function given by the expectation value
$H_Q(\psi)=\langle\psi,\hat{H}\psi\rangle$, seen as a function on the
Hilbert space. This follows easily if we choose the orthonormal basis
above to be given by eigenstates of the Hamiltonian operator,
$\hat{H}|\psi_j\rangle=E_j|\psi_j\rangle$. Then,
$H_Q(c_j)=\sum_jE_j|c_j|^2= \sum_jE_j(({\rm Re} c_j)^2+({\rm Im}
c_j)^2)$ and we have equations of motion
\[
 \frac{\md}{\md t}{\rm Re} c_j = \{{\rm Re} c_j,H_Q\}=
\frac{E_j}{\hbar}{\rm Im} c_j \quad,\quad
 \frac{\md}{\md t}{\rm Im} c_j = \{{\rm Im} c_j,H_Q\}=
-\frac{E_j}{\hbar}{\rm Re} c_j
\]
or $\dot{c}_j=-i\hbar^{-1}E_j c_j$ with solution
$c_j(t)=\exp(-i\hbar^{-1}E_jt)$ which is equivalent to the solutions
of the Schr\"odinger equation.

This geometrical picture of quantum mechanics
\cite{GeomQuantMech} has been used for
semiclassical definitions in \cite{Schilling,Josh}, and in
\cite{EffAc} for developing the point of view of effective systems
used here.\footnote{Some of these ideas look related to ingredients of
Ehrenfest theorems or the WKB approximation. However, the geometrical
formulation allows much tighter control over all the correction terms
to classical behavior. For an-harmonic oscillators discussed later, for
instance, the WKB approximation agrees with effective action results
only to first order in the anharmonicity parameter \cite{EffAcWKB}.}
There are thus truly infinitely many quantum degrees of freedom,
although their role as higher derivatives of classical variables or
something else can only be found after studying dynamical equations.

In fact, the quantum variables are dynamical and in general back-react
on the classical variables: the $G^{a,n}$ change if a wave packet
spreads and deforms. For a quadratic Hamiltonian in canonical
variables, we can simply use relations such as
$\langle\hat{q}^2\rangle=\langle\hat{q}\rangle^2+G^{0,2}$ which imply
that terms containing the quantum variables are just added to the
classical Hamiltonian. This can give zero point energy contributions,
but since no products between classical and quantum variables occur,
there are no coupling terms.  If the classical Hamiltonian is not
quadratic in the canonical variables, on the other hand, coupling
terms between the quantum and classical variables occur by expanding
the expectation value of the Hamiltonian operator in a general
state. To make this explicit, we write the quantum Hamiltonian as
\begin{eqnarray}
 H_Q &=&\langle H(\hat{q},\hat{p})\rangle_{\rm Weyl}=\langle
H(q+(\hat{q}-q),p+(\hat{p}-p))\rangle_{\rm Weyl}\nonumber\\
&=&\sum_{n=0}^{\infty} \sum_{a=0}^n
\frac{1}{n!}\left(\begin{array}{c} n \\
a\end{array}\right)
\frac{\partial^n
H(q,p)}{\partial p^a\partial q^{n-a}}G^{a,n}
\end{eqnarray}
where from now on $q=\langle\hat{q}\rangle$ and
$p=\langle\hat{p}\rangle$.  Since $G^{a,0}=1$ and $G^{a,1}=0$ by
definition, the quantum variables appear starting at second order
$G^{a,2}$ with coefficients proportional to $\partial^2
H(q,p)/\partial p^a\partial q^{2-a}$. These are constants for a
quadratic potential, where all higher coefficients vanish, but do
depend on the classical variables for Hamiltonians with non-harmonic
potentials or non-standard kinetic terms. In the latter cases, thus,
coupling terms between classical and quantum variables arise.

For the dynamical behavior of classical and quantum variables we need to
compute the Hamiltonian equations of motion
\begin{equation}
 \dot{q}=\{q,H_Q\}\quad,\quad
 \dot{p}=\{p,H_Q\}\quad,\quad
 \dot{G}^{a,n}=\{G^{a,n},H_Q\}\,.
\end{equation}
For this, we need to know the symplectic structure \cite{EffAc} which
is given by the classical Poisson brackets $\{q,p\} =1$ together with 
$\{q,G^{a,n}\}=0=\{p,G^{a,n}\}$
and
\begin{eqnarray}\nonumber&\left\{G^{a,n},G^{b,m}\right\}=
\sum_r \left[(_2^{\underline{\hbar}})^{2r}
K[a,b,m,n,r]G^{a+b-2r-1,m+n-4r-2}\right]& \\&
-b(n-a)G^{a,n-1}G^{b-1,m-1}+a(m-b)G^{b,m-1}G^{a-1,n-1}&
\end{eqnarray}
where \begin{equation}K[a,b,m,n,r]={\sum_{0\leq f\leq
2r+1}}(-)^{r+f}(f!(2r+1-f)!)^{-1}\left(^{a}_{f}\right)\left(^{n-a}_{2r+1-f}\right)
\left(^{b}_{f}\right)\left(^{m-b}_{2r+1-f}\right) .\end{equation}

If we rescale our quantum variables by $\tilde
G^{a,n}=\hbar^{-n/2}(m\omega)^{n/2-a}G^{a,n}$ to make them
dimensionless and compute the equations for an an-harmonic oscillator
with classical Hamiltonian $H=
\frac{1}{2m}p^2+\frac{1}{2}m\omega^2q^2+U(q)$, we obtain the quantum
Hamiltonian
\begin{equation} \label{HQexpand}
H_Q=\frac{1}{2m}p^2+\frac{1}{2}m\omega^2q^2+U(q)
+\frac{\hbar\omega}{2}(\tilde{G}^{0,2}+\tilde{G}^{2,2})+\sum_n\frac{1}{n!}
\left(\frac{\hbar}{m\omega}\right)^{n/2}U^{(n)}(q)\tilde{G}^{0,n}
\end{equation}
as an expansion in $\hbar$ which clearly shows the coupling terms
between classical and quantum variables coming from a non-quadratic
potential. With the above Poisson brackets, $H_Q$ generates equations of
motion
\begin{eqnarray}
\dot{q}&=& \frac{p}{m}\label{eom}\\
\dot{p}&=&-m\omega^2q -U'(q)-\sum_n\frac{1}{n!}\left(
\frac{\hbar}{m\omega}\right)^{n/2}U^{(n+1)}(q)\tilde{G}^{0,n}\\
\dot{\tilde{G}}{}^{a,n}&=&-a\omega
\tilde{G}^{a-1,n}+(n-a)\omega \tilde{G}^{a+1,n}
-a\frac{U''(q)}{m\omega}\tilde{G}^{a-1,n}\\
\nonumber&&+
\frac{\sqrt{\hbar}aU'''(q)}{2(m\omega)^{\frac{3}{2}}}\tilde{G}^{a-1,n-1}
\tilde{G}^{0,2}
+\frac{\hbar aU^{''''}(q)}{3!(m\omega)^2}\tilde{G}^{a-1,n-1}\tilde{G}^{0,3}\\
\nonumber&&
-\frac{a}{2}\left(
\frac{\sqrt{\hbar}U'''(q)}{(m\omega)^{\frac{3}{2}}}
\tilde{G}^{a-1,n+1}+\frac{\hbar
U^{''''}(q)}{3(m\omega)^2}\tilde{G}^{a-1,n+2}\right)\\
\nonumber&&
+\frac{a(a-1)(a-2)}{24}\left(
\frac{\sqrt{\hbar}U'''(q)}{(m\omega)^{\frac{3}{2}}}G^{a-3,n-3}+\frac{\hbar
U^{''''}(q)}{(m\omega)^2}\tilde{G}^{a-3,n-2}\right)+\cdots\,.
\end{eqnarray}
These infinitely many coupled ordinary differential equations are
equivalent to the Schr\"odinger equation and in general not easier to
solve. However, this set of equations is much more suitable for
splitting the dynamics in classical effective equations and equations
for quantum degrees of freedom. We also note that such a system of
infinitely many differential equations requires infinitely many
parameters as initial conditions. Although they are not completely
arbitrary, as the quantum variables have to satisfy uncertainty
relations such as $G^{0,2}G^{2,2}\geq\frac{\hbar^2}{4}+(G^{1,2})^2$,
fixing these values will turn out to be one of the crucial parts of
deriving effective equations.

\subsection{Example: Harmonic oscillator}

The Hamiltonian of the harmonic oscillator is quadratic in the
canonical variables and there are thus no coupling terms between
classical and quantum variables. This conforms with the well-known
fact that there are dynamical coherent states for the harmonic
oscillator which do neither spread nor deform while following the
classical trajectories exactly. At the quantum level, we have
Hamiltonian equations of motion
\begin{eqnarray}
\dot{p}&=&\{p,H_Q\}=-m\omega^2 q\\
\dot{q}&=&\{q,H_Q\}=\frac{1}{m} p\\
\dot{G}^{a,n}&=&\{{G}^{a,n},H_Q\}=\frac{1}{m}(n-a)G^{a+1,n}-m\omega^2
aG^{a-1,n}
\end{eqnarray}
which indeed decouples to an infinite set of finitely many coupled
differential equations. Moreover, one can see that constant solutions
for the quantum variables exist,
\begin{equation}\label{freemoments}
 G^{a,n}=2^{-n}\hbar^{1/2}(m\omega)^{a-n/2}
\frac{a!}{(a/2)!}\frac{(n-a)!}{((n-a)/2)!}\,,
\end{equation}
saturating the uncertainty relations. These are exactly the solutions
corresponding to coherent states.  Since quantum variables do not
appear in the above equations of motion for classical variables, there
is no need to introduce new effective equations. Indeed, effective
actions for ``free'' theories such as the harmonic oscillator are
always identical to the classical action. Nevertheless, even in this
simple example we can already see the generality of the effective
equation scheme discussed here: We can just as well choose
non-constant solutions for $G^{a,n}$ which then change cyclically
along the classical orbit. This describes wave packets which deform
periodically while following classical peak positions, a behavior
which can be interpreted as semiclassical, just as the case of
constant $G^{a,n}$, provided that the $G^{a,n}$ do not change too
rapidly.

\subsection{An-harmonic oscillator}

The situation is more interesting for non-quadratic classical
Hamiltonians because for them coupling terms between classical and
quantum variables appear. All infinitely many differential equations
for the classical and quantum variables are then coupled and $(q,p)$
are affected by the motion of $G^{a,n}$ in non-trivial ways. This
describes the back-reaction of spreading and deformations of the wave
packet on their peak positions.

For practical purposes, this set of infinitely many coupled equations
must be truncated to a finite set in suitable approximations, such as
the adiabatic approximation in quantum variables.  In this
approximation, equations for $G^{a,n}$ can be solved perturbatively in
$\hbar$ and in the adiabatic expansion. The latter is an expansion in
a parameter $\lambda$ formally introduced in the calculation, but in
the end set to $\lambda=1$. Derivatives with respect to time in
equations of motion are first re-scaled as $\frac{\md}{\md
t}\rightarrow\lambda\frac{\md}{\md t}$. Moreover, slowly changing
variables for which the adiabatic approximation is done are expanded
in a series in $\lambda$ as well. We expand only the quantum variables
$G^{a,n}=\sum_eG^{a,n}_e\lambda^e$ in this manner, meaning that their
change in time is adiabatic, but keep the evolution of classical
variables free. After inserting this in the equations of motion and
expanding in $\lambda$, one obtains equations for all coefficients
$G^{a,n}_e$ at different orders of the adiabatic
expansion.\footnote{For more details on the following calculations and
underlying definitions, see \cite{EffAc}.} The equations of motion
$\dot{G}^{a,n}=\{G^{a,n},H_Q\}$ for quantum variables then imply
$\dot{G}_{e-1}^{a,n}=\{G_e^{a,n},H_Q\}$.  In addition to the adiabatic
approximation there is also a semiclassical expansion in powers of
$\hbar$. To obtain effective equations to the order relevant for
(\ref{EffAc}), it turns out that one has to calculate the first order in
$\hbar$ and go to second order in $\lambda$ for $G^{a,2}$.

We can now use the Poisson relations
\[
 \{G^{a,n},G^{0,2}\} =-2aG^{a-1,n} \quad \mbox{ and } \quad
 \{G^{a,n},G^{2,2}\} =2(n-a)G^{a+1,n}
\]
giving equations
\begin{equation} \label{zerol}
 0=\{G^{a,n}_0,H_Q\}=\omega\left((n-a)G_0^{a+1,n}-a
 \left(1+\frac{U''(q)}{m\omega^2}\right)G_0^{a-1,n}\right)
\end{equation}
at zeroth order in $\lambda$ used for all $n$,
\begin{equation}\label{firstl}
 \dot{G}_0^{a,n}=\{G_1^{a,n},H_Q\}=
 \omega\left((n-a)G_1^{a+1,n}-a\left(1+\frac{U''(q)}{m\omega^2}\right)
 G_1^{a-1,n}\right)
\end{equation}
to first order in $\lambda$ used for all $n$, and
\begin{equation} \label{secondl}
 G^{2,2}_2-\left(1+\frac{U''(q)}{m\omega^2}\right)
 G^{0,2}_2=\frac{1}{\omega}\dot{G}_1^{0,2}=\frac{1}{2\omega^2}\ddot G^{0,2}_0
\end{equation}
to second order in $\lambda$ at $n=2$.

The general solution of (\ref{zerol}) is
\[
 G_0^{a,n}=\left(\begin{array}{c}n/2\\ a/2\end{array}\right)
\left(\begin{array}{c}n\\ a\end{array}\right)^{-1}\left(
 1+\frac{U''(q)}{m\omega^2}\right)^{a/2}G^{0,n}_0
\]
for even $n$ and $a$ and zero whenever $a$ and/or $n$ are odd. This
still leaves the value of $G_0^{0,n}$ free, which will be fixed
shortly. The first order equation (\ref{firstl}) then implies
\begin{eqnarray*}
&& 
 \frac{1}{\omega}\sum_{a} \left(\begin{array}{c}n/2\\ a/2\end{array}\right)
 \left(1+\frac{U''(q)}
 {m\omega^2}\right)^{(n-a)/2}\dot{G}_0^{a,n}\\
 &=&
 \sum_{a\:\: {\rm even}}\left(\begin{array}{c}n/2\\ a/2\end{array}\right)
 \left(1+\frac{U''(q)}
 {m\omega^2}\right)^{(n-a)/2}\left(
 (n-a)G_1^{a+1,n}-a\left(1+\frac{U''(q)}{m\omega^2}\right)
 G_1^{a-1,n}\right)\\
 &=&0
\end{eqnarray*}
which can be seen by shifting $a\to a-2$ in the first term of the
right hand side.  This imposes a constraint on $G_0^{0,n}$ solved by
$G^{0,n}_0=C_n(1+\frac{U''(q)}{m\omega^2})^{-n/4}$.  The remaining
constants $C_n$ can be fixed to $C_n=\frac{n!}{2^n(n/2)!}$ by
requiring that the limit $U(q)\rightarrow 0$ reproduces the quantum
variables (\ref{freemoments}) of coherent states of the harmonic
oscillator. This means that we require the perturbative vacuum of the
quantum theory to be reproduced in the effective system. Therefore,
\[
 G_0^{a,n}=\frac{(n-a)!a!}{2^n((n-a)/2)!(a/2)!}
 \left(1+\frac{U''(q)}{m\omega^2}\right)^{\frac{2a-n}{4}}\,.
\]

{}From the first order corrections $G_1^{a,n}$, we will only need
solutions for $n=2$ which follow directly from (\ref{firstl}) with
$a=0$ as $G^{1,2}_1=\frac{1}{2\omega}\dot G^{0,2}_0$. The second order
equation (\ref{secondl}) again leaves free parameters in the general
solution to be fixed by the next, third order:
$\left(1+U''(q)/(m\omega^2)\right)\dot G^{0,2}_2+\dot G^{2,2}_2=0$ as
before. From this, the solution to the system is
\begin{eqnarray*}
 G_2^{0,2}&=&-\frac{2}{\omega^2} (G^{0,2}_0)^{\frac{5}{2}}(
 (G^{0,2}_0)^{\frac{1}{2}})\ddot{}\\
 &=&\frac{\left(
 1+\frac{U''(q)}{m\omega^2}\right)^{-\frac{7}{2}}}
 {4\omega^2}\left(\left(1+\frac{U''(q)}{m\omega^2}\right)\frac{U'''(q)\ddot
 q+U''''(q)\dot q^2}{4m\omega^2}-5\left(\frac{U'''(q)\dot
 q}{4m\omega^2}\right)^2\right)\,.
\end{eqnarray*}

Finally, putting our approximate expressions for the quantum variables
back into the equations
\begin{eqnarray}
\dot{q}&=& m^{-1}p\\
\dot{p}&=&-m\omega^2q -U'(q)-\sum_n\frac{1}{n!}(m^{-1}\omega^{-1}
\hbar)^{n/2}U^{(n+1)}(q)G^{0,n}
\end{eqnarray}
for the classical variables and writing them as a second order
equation for $q$, we obtain
\begin{eqnarray}
&&\nonumber\left(m+\frac{\lambda^2\hbar U'''(q)^2}{32m^2\omega^5\left(
 1+\frac{U''(q)}{m\omega^2}\right)
 ^{\frac{5}{2}}}\right)\ddot q\\
\nonumber&&+\frac{\lambda^2\hbar\dot
 q^2\left(4m\omega^2U'''(q)U''''(q)\left(1+\frac{U''(q)}{m\omega^2}\right)-
 5U'''(q)^3\right)}
 {128m^3\omega^7\left(1+\frac{U''(q)}{m\omega^2}\right)^{\frac{7}{2}}}\\
&&+m\omega^2q+U'(q)+\frac{\hbar
U'''(q)}{4m\omega\left(1+\frac{U''(q)}{m\omega^2}\right)^{\frac{1}{2}}}=0\,.
\end{eqnarray}
After setting $\lambda=1$, we finally have our effective equations to
first order in $\hbar$, which agree with the equations of motion
determined by (\ref{EffAc}). Thus, the methods described here provide
a generalized derivation of effective equations, and in particular a
Hamiltonian picture.

\subsection{General effective systems}

Since the main difference for dynamical purposes between classical and
quantum mechanical systems lies in the infinite dimensionality of a
quantum system compared to the finite dimensionality of a classical
mechanical one, the main aim of any effective approximation is a
truncation of the quantum dynamics to an effective one on a finite
dimensional subspace. Moreover, almost all of the infinitely many
integration constants, corresponding to the shape of an initial
quantum state, required for the full quantum system have to be fixed
by some means. In the above procedure for an an-harmonic oscillator,
the adiabatic approximation together with an $\hbar$-expansion
required only a finite number of $G^{a,n}$ at each order, resulting in
a truncation of the quantum system to an effective one.  The
integration constants such as $C_n$ were fixed by relating them to
properties of harmonic oscillator coherent states.  But even in this
case, other choices are possible and will result in different
effective equations describing states which do not saturate
uncertainty relations or are squeezed. This shows that effective
equations in general terms will never be unique, in contrast to the
low energy effective action obtained by perturbing around the vacuum
state, but will occur in parameterized form to describe different
physically relevant sectors. The low energy effective action is not
suitable for any such situation that may arise because one cannot
describe arbitrary states as perturbations around a ground state.

The example of an-harmonic oscillators also elucidates the role of
higher derivatives and their potential relation to quantum degrees of
freedom. In the geometrical picture, there is a consistent way to
introduce additional degrees of freedom by keeping some of the
$G^{a,n}$ as independent variables whose solutions are not inserted
into the equations of classical variables. Only the remaining quantum
variables are then solved approximately and inserted in equations for
classical variables as well as those of the quantum variables we
kept. This gives a higher dimensional effective system with new,
non-classical degrees of freedom. While it may then be possible in
some cases to view those additional degrees of freedom as higher
derivatives of the classical ones, after making explicit use of
equations of motion, this will not be the same as higher derivatives
in a higher derivative Lagrangian. In the latter case, higher
derivatives as degrees of freedom are only implicit as discussed
earlier, while keeping additional $G^{a,n}$ as true independent
degrees of freedom is consistent in the perturbation scheme. Implicit
effects of higher derivative terms in a low energy effective action
correspond, rather, to using higher orders in an adiabatic
approximation of the quantum variables following the same procedure as
before, i.e.\ using only $q$ and $p$ in effective equations.

An example where one has to use some quantum variables as independent
degrees of freedom is the free particle. The adiabatic approximation
is not consistent in this case, as one can see from the fact that the
effective action (\ref{EffAc}) diverges for a potential
$U(q)=-\frac{1}{2}m\omega^2q^2$ which cancels the harmonic
contribution exactly. This is analogous to infrared problems in the
massless limit of quantum field theory. In this case, the spreading of
wave packets is too strong to allow an adiabatic approximation, but
keeping the spread as one of the independent quantum variables allows
a well-defined effective formulation \cite{EffAc}. A more interesting
example is cosmological structure formation where keeping quantum
variables of order two in effective equations shows how quantum
fluctuations seed classical metric perturbations \cite{StructureGen}.

\subsection{Comparison}

For free theories such as the harmonic oscillator there is no
difference between the standard low energy effective action and the
general methods described here, because quantum variables do not
couple to classical variables. Thus, although one can choose
non-vacuum states to expand around and thereby obtain different
dynamics for the quantum variables, this does not back-react on the
dynamics of classical variables. There are other common properties
such as the fact that low energy effective actions for non-quadratic
theories arise from non-local objects in time, such that no truncation
to finite order in higher derivatives can give exact results, which is
related to the fact that there are infinitely many variables in a
quantum theory and thus infinitely many coupled equations. In the same
way as a local approximation of the low energy effective action can be
done by performing a derivative expansion, the infinitely many quantum
variables of general effective systems can be cut down to a finite
number in an adiabatic approximation. Non-local features of the
low-energy effective action occur through higher derivative terms,
suggesting additional degrees of freedom which may be seen in analogy
to the independent quantum variables. Here, both pictures are starting
to become different because higher derivative actions in a
perturbative formulation are more subtle, as discussed before, and do
not obviously have additional degrees of freedom, while quantum
variables are true degrees of freedom of a quantum theory. There is
also a difference in the interpretation of variables entering
effective equations: In the usual picture of low-energy effective
actions, ``classical'' variables are related to non-diagonal matrix
elements of operators which are not guaranteed to be real
\cite{VariationalEffAc}. In the general scheme of effective systems, on the
other hand, classical variables are expectation values of the basic
operators and thus always real.

Further differences refer to the general form of effective equations
and their uniqueness. While the low energy effective action is by
definition obtained by perturbing around the vacuum state, one can
derive effective systems more generally for any state. Thus, while
there are no free parameters in the low energy effective action which
appears to be unique, initial conditions of the quantum variables
describing spread and deformations of the initial state appear in
general effective equations. In a situation where one has a unique way
to fix the state, such as the harmonic oscillator ground state, one
reproduces low energy results by the more general method. This
comparison is summarized in Tab.~\ref{Comp}.

\begin{table}[ht]
{\begin{tabular}{@{}cc@{}} \hline
low energy & general\\\hline\hline
\multicolumn{2}{c}{free theories (harmonic oscillator) unchanged}\\\hline
{\em non-local} & {\em $\infty$-many coupled} equations\\\hline
derivative expansion & adiabatic approximation\\\hline
{\em higher derivatives} & {\em independent quantum variables}\\\hline
``classical'' variables related to& {\em expectation values} in\\
{\em non-diagonal matrix elements} & {\em dynamical coherent states}\\\hline
expansion {\em around free theory} & expansion possible around\\
 & {\em any state}, such as squeezed ones\\\hline
``unique,'' {\em free of parameters} & free parameters from {\em initial
conditions}\\
&  {\em of $G^{a,n}$} specify state to expand around\\\hline
\multicolumn{2}{c}{usual
effective action (\ref{EffAc}) for expansion around the harmonic
oscillator vacuum}\\ \hline
\end{tabular}}
\caption{Comparison between properties of low energy effective actions and
general effective systems. \label{Comp}}\end{table}

\section{Implications for quantum gravity}
\label{Impl}

By embedding the standard procedure to arrive at low energy effective
actions into a general scheme, we have seen that more general
effective descriptions cannot be unique. They rather depend on
parameters corresponding to the choice of a state to expand around. In
low energy effective actions, this state is chosen to be the
perturbative vacuum state which fixes the parameters, but this is not
available for more general situations. The low energy effective action
is in fact what is relevant for low energies and only small
excitations out of the vacuum. But for applications in gravity, unless
one is dealing with graviton scattering, the applicability of a low
energy effective action is questionable. This is in particular true
for quantum cosmology based on effective actions because the whole
universe is far from a gravitational vacuum state, however this may be
defined. For instance, in background dependent quantizations of
gravity, the background metric is used to define the perturbative
vacuum, resulting in higher curvature low energy effective actions. It
is then sometimes tempting to use these effective actions for a
general metric and study implications even in strong curvature regimes
of black holes or cosmology far away from the original
background. However, this is clearly outside the allowed range of
validity of the approximation.

A quantum cosmological situation is not expected to be described by
expanding around the vacuum state of a free field theory; an effective
description should rather be obtained by expanding around a suitable
initial state describing a semiclassical universe at large scales but
being far from a vacuum state. Unlike the vacuum, such states are not
unique but depend on several parameters. Then, also effective
equations obtained by such an approximation depend on the same
parameters: they appear as initial values for the quantum variables
$G^{a,n}$ which, when solutions are inserted in the equations for
classical variables, also enter the effective equations. Thus, a
general effective description relevant for most purposes of gravity or
cosmology cannot be as unique as an effective action obtained by
perturbing around the vacuum. Parameters that will appear in general
are not simply ambiguities of the formalism, but they have physical
relevance as they describe properties such as the spread of an initial
state. The dynamics will in general depend on the spread such that it
must enter the effective equations. The choice of parameters thus can
be fixed by physical considerations, just as the vacuum state chosen
in low energy effective actions determines the initial spread and
deformations.

This is the situation naturally encountered in background independent
quantum theories of gravity where it is anyway more difficult to
define a perturbative vacuum state. One would thus use a class of
semiclassical states, parameterized in some way if no distinguished
state is known, to compute expectation values of Hamiltonians. These
parameters then appear in effective equations as they do even in
mechanical systems. Some of those parameters also occur in effective
actions derived from background dependent quantizations since one can
sometimes use different backgrounds to perturb around. However, there
are other, truly quantum parameters which are not taken into account
by just changing the background, but which may be important
physically.

Finally, while the vacuum state is Lorentz invariant and the
corresponding effective action is covariant, a general state will not
be preserved by a Lorentz transformation. This is in particular true
for a semiclassical state corresponding to an expanding
cosmology. Thus, effective equations relevant for quantum gravity or
cosmology should not be expected to be Lorentz covariant. Performing a
Lorentz transformation maps the chosen state to a different one,
thereby changing the parameters appearing in effective
equations. These equations, then, also change to a different
set. (Sometimes, in particular when parameters come from different
metric backgrounds, one can view a whole class of effective systems as
one system invariant under Lorentz transformations, for which also
background parameters change. But all these effective systems and the
background parameters really come from different states which are then
being transformed into each other.) Even if the original theory of
quantum gravity for which an effective system is considered were
covariant, the effective equations will not be. This comes about
because one is forced to treat all the degrees of freedom entering the
quantum system differently, by formulating effective equations for
some of them but keeping the rest only in a parameterized way. This
cannot be avoided because the whole point of an effective description
is to cut down the infinite number of quantum variables to a finite
set. Lorentz transformations at the quantum level, in general, mix all
these parameters and cannot be constrained to act on just the
variables kept for an effective system. It is then impossible to draw
conclusions concerning Lorentz invariance or violation based on just
one set of effective equations, but a wider view based on more general
properties of effective systems can give indications for symmetries of
a quantum theory of gravity.

\section*{Acknowledgments}

We are grateful to Abhay Ashtekar and J\"urg K\"appeli for
discussions.  MB thanks the organizers of the 42nd Karpacz Winter
School of Theoretical Physics ``Current Mathematical Topics in
Gravitation and Cosmology,'' Ladek, Poland, February 6--11, 2006 and
in particular Andrzej Borowiec for an invitation to give a lecture
course on which this text is based.

\appendix

\section{Loop quantum gravity and cosmology}

In these appendices we summarize the situation in the main current
candidates for quantum theories of gravity, loop quantum gravity and
string theory, from the point of view of effective actions and higher
curvature terms. This presents one example each for the two different
classes of background independent and background dependent
quantizations.
 We provide more details on loop quantum gravity since
this is the lesser known framework (which is also subject to the
common misconception that effective correction terms would not arise
here) but closer to the authors' expertise, and certainly not because
there would be more results on effective actions from loop quantum
gravity.

Loop quantum gravity is based on a canonical quantization such that
space and time coordinates are treated differently. This is relevant
for the appearance of higher derivative terms because time-derivatives
are replaced by momenta while space-derivatives are
retained. Moreover, the loop quantization provides a discrete spatial
structure such that difference operators rather than differential ones
occur. It is thus easy to see higher spatial derivatives emerging for
effective field equations, but for higher time derivatives one has to
do more involved calculations using quantum variables as described
before. Based on the presence of higher spatial derivatives but no
higher time derivatives, possible Lorentz violations have been
suggested \cite{GRB}, but as mentioned in
Sec.~\ref{Impl} a final verdict has to await a more detailed
analysis. This requires also higher time derivative corrections which
have not been derived yet for any model of loop quantum gravity, but
they can certainly arise along the lines of quantum variables
described in the main text. To see why higher spatial derivatives must
occur, we present a brief, mostly self-contained introduction to loop
quantum gravity. More detailed reviews of these aspects are, e.g.,
\cite{Rev}.

\subsection{Canonical quantization}

Rather than using ADM variables\footnote{The canonical ADM variables
\cite{ADM} for general relativity are defined in terms of the spatial
metric $q_{ab}$ and extrinsic curvature $K_{ab}=\frac{1}{2N}({\cal
L}_tq_{ab}-2D_{(a}N_{b)})$, where the time function $t$ determines
spatial slices $\Sigma_t\colon t={\rm const}$ and time is measured
along a time evolution vector field $t^a=Nn^a+N^a$ decomposed into
lapse function $N$ and shift vector $N^a$ using the unit normal $n^a$
to $\Sigma$.}  as in a Wheeler--DeWitt quantization, loop quantum
gravity uses Ashtekar variables \cite{AshVar,AshVarReell} where the
field equations are easier to handle.  The spatial metric is then
replaced by a densitized triad $E^a_i$, such that
$E^a_iE^b_i=q^{ab}\det q$, which also determines the spin connection
$\Gamma_a^i$ compatible with $E^a_i$. The spin connection, in turn,
together with extrinsic curvature components determines the connection
$A_a^i=\Gamma_a^i+\gamma K_a^i$ (with Barbero--Immirzi parameter
$0<\gamma\in {\mathbb R}$ \cite{AshVarReell,Immirzi}) which is
canonically conjugate to the densitized triad.

The Hamiltonian consists only of constraints due to general
covariance, $H=G[\Lambda]+ D[N^a]+H[N]=0$, with the Gauss constraint
$G[\Lambda]$ (generating triad rotations), the diffeomorphism
constraint $D[N^a]$ (generating spatial diffeomorphisms) and the
Hamiltonian constraint
\begin{eqnarray}\label{HamConstr}
 H[N] &=& \frac{1}{16\pi G} \int_{\Sigma} \mathrm{d}^3x N\left|\det
   E\right|^{-1/2}
 \Bigl(\epsilon_{ijk}F_{ab}^iE^a_jE^b_k\\
 && -2(1+\gamma^{-2})
 (A_a^i-\Gamma_a^i)(A_b^j-\Gamma_b^j)E^{[a}_iE^{b]}_j\Bigr)\nonumber
\end{eqnarray}
which is the most complicated of the constraints.
In this expression, $F_{ab}^i$ are curvature components of the
connection $A_a^i$ and thus quadratic in the canonical variables, but
the coefficients $\Gamma_a^i$ are complicated functions of
$E^a_i$. All these components contain spatial derivatives while time
derivatives are replaced by momenta.

To illustrate these variables we consider isotropy as an example. The
fields can then be written as $E^a_i=p\delta^a_i$ and
$A_a^i=c\delta_a^i$ with only two remaining canonical variables
$|p|={\textstyle\frac{1}{4}}a^2$ and
$c={\textstyle\frac{1}{2}}(k+\gamma\dot{a})$ where $a$ is the scale
factor of a Friedmann--Robertson--Walker metric, $\dot{a}$ its
derivative in proper time, and $k=0$ or $k=1$ the curvature parameter
which enters through the spin connection. (For negative spatial
curvature $k=-1$ the expression for the connection is not of the above
diagonal form \cite{Kevin}.) 

The isotropic Hamiltonian constraint simplifies compared to the full
expression and is given by
\begin{equation} \label{Hamiso}
  H=-\frac{3}{8\pi
   G}(\gamma^{-2}(c-k/2)^2+k^2/4)\sqrt{|p|}+H_{\mathrm{matter}}(p)=0
\end{equation}
with some matter Hamiltonian $H_{\rm matter}$. Using the
transformation of variables above, this can easily be checked to
reduce to the Friedmann equation while it generates as Hamiltonian
equations of motion the Raychaudhuri equation by $\dot{p}=\{p,H\}$,
$\dot{c}=\{c,H\}$ as well as matter evolution equations if matter
fields are present.

To proceed with the full theory, we then have to see how to represent
the canonical variables as an operator algebra on a Hilbert space such
that Poisson relations become commutator relations. This requires one
to smear fields by integrating them over suitable regions, but in a
background independent quantization should be done in such a way that
no background metric is introduced.

\subsection{Holonomies and fluxes}

At this point, we encounter one of the main advantages of Ashtekar
variables: they allow a natural smearing of basic fields
$(A_a^i,E^b_j)$ to linear objects without introducing a background,
while still leading to a well-defined algebra. This comes about
because we can naturally integrate a connection along a curve, also
taking the path ordered exponential to have good gauge transformation
properties under local SU(2), to give holonomies $h_e(A)={\cal
P}\exp\int_e\tau_i A_a^i\dot{e}^a{\rm d}t$ for an arbitrary curve $e$
in the spatial manifold $\Sigma$. Similarly, a densitized vector field
can naturally be integrated over 2-surfaces $S$ to obtain fluxes
$F_S(E)=\int_S \tau^i E^a_in_a{\rm d}^2y$. This integration in one
plus two dimensions, without introducing any background
measure,\footnote{The edges $e$ and surfaces $S$ are not fixed but
appear as labels, just as points $x$ appear as labels in usual field
formulations. Topological and differential background structures still
need to be chosen to define the basic objects, but this does not
prevent background independence to be realized as in classical general
relativity. Also there, one has to choose a topology and differential
structure before formulating the field equations, but one does not
split the metric into a background plus fields on that
background. Background independence in loop quantum gravity is to be
understood in the same sense.} turns out to remove all delta functions
in the classical Poisson relations and thus results in a well-defined
algebra to be represented on a Hilbert space.

Such a representation is the basis for a background independent
quantization \cite{LoopRep}. One also has to require that the
representation one is using is covariant under spatial diffeomorphisms
since these transformations have to be removed as gauge. Under this
condition, an irreducible, cyclic representation of the holonomy-flux
algebra is then uniquely determined \cite{LOST}. Thus, all states can
be obtained from operators acting on a ``ground state'' in which no
geometry at all is excited.  One can construct all states in the
connection representation, i.e.\ as functionals of the connection, by
using holonomies as multiplication operators. The ``ground state'' is
just a constant on the space of connections, and by multiplication with
holonomies one ``creates'' dependence on the connection along
edges. This results in spin network states \cite{RS:Spinnet} of the
form
\[
 T_{g,j,C}(A)=\prod_{v\in g} C_v\cdot \prod_{e\in g}
  \rho_{j_e}(h_e(A))
\]
which are labeled by an oriented graph $g$, irreducible SU(2)
representations $j$ on its edges and gauge invariant contraction
matrices $C$ in its vertices. An example is the Wilson loop in the
fundamental representation,
\[
 W_g(A) = \epsilon_{AC}\epsilon^{BD}\cdot (h_1(A))^A_B
(h_2(A))^C_D
 = (h_1(A))^A_B (h_2(A)^{-1})^B_A = {\rm tr}(h_1(A)
h_2(A)^{-1})\,,
\]
for a loop seen as two edges from one vertex to another one oriented
in the same way.  This gives two holonomies $h_1$ and $h_2$ between
the vertices on which the $\epsilon$ tensors are contraction
matrices. In this way, one automatically obtains the trace of a closed
holonomy which is gauge invariant, while using other contractions
would not give an invariant function. More generally, one can
construct gauge invariant spin network states with intersection points
and form a basis of all states on connections.

\subsection{Discrete geometry}

Fluxes, being conjugate to holonomies, become derivative operators on
spin network states acting as
\[
 \hat{F}_S f_g = -8\pi i\gamma G\hbar \int_S {\rm d}^2y\tau^i n_a
 \frac{\delta f_g(h(A))}{\delta A_a^i(y)}
 =-8\pi i\gamma\ell_{\rm P}^2
\sum_{e\in g}\int_S {\rm d}^2y\tau^i n_a
\frac{\delta h_e}{\delta
    A_a^i(y)}\frac{\md f_g(h)}{\md
    h_e}\,.
\]
Here, the Planck length $\ell_{\rm P}=\sqrt{G\hbar}$ arises
automatically.  There are non-zero contributions only if the surface
$S$ in the flux intersects edges of the graph $g$ of the state, and
individual contributions from intersection points are determined by
``angular momentum operators'' (su(2) derivatives) acting on
holonomies. Since such operators have discrete spectra, one sees
that fluxes and thus spatial geometry are discrete. Operators such as
area and volume can be constructed from fluxes alone and inherit the
discreteness of spectra \cite{AreaVol}. Zero is always an
eigenvalue contained in the spectra, usually of high degeneracy,
corresponding to space which is ``empty'' even of geometry.

While this property indicates discreteness of quantum geometry at
least spatially and at the kinematical level, the fact that zero is an
eigenvalue in the discrete part of the spectrum also spells trouble:
it means that the volume operator, or any of its local contributions
corresponding to $\sqrt{\det q}$, does not have a densely defined
inverse.  However, inverse powers of the determinant of the metric are
required for matter Hamiltonians as well as the Hamiltonian
constraint. For a scalar field, for instance, we have the classical
matter Hamiltonian
\begin{equation}
  H_{\phi}=\int\mathrm{d}^3x\left( \frac{1}{2}
   \frac{p_{\phi}^2+E^a_iE^b_i\partial_a\phi
     \partial_b\phi}{\sqrt{|\det E^c_j|}}+\sqrt{|\det
     E^c_j|}V(\phi)\right)
\end{equation}
where not only the field variables are to be quantized but also metric
components, in particular the inverse determinant, in quantum gravity.

Fortunately, this problem can be solved using identities such as
\cite{QSDI}
\begin{equation}
 \left\{A_a^i,\int\sqrt{|\det E|}\mathrm{d}^3x\right\}= 2\pi\gamma G
 \epsilon^{ijk}\epsilon_{abc} \frac{E^b_jE^c_k}{\sqrt{|\det E|}}
\end{equation}
which allow one to express an inverse power of densitized triad
components by a Poisson bracket between connection components and only
positive powers of the triad components. The connection components
$A_a^i$, furthermore, can be approximated by holonomies, which is
necessary because only holonomies are represented on the Hilbert space.
Inserting appropriate holonomies and the volume operator, and
replacing the Poisson bracket by $(i\hbar)^{-1}$ times a commutator
results in a well-defined operator which has the correct classical
limit corresponding to inverse powers of densitized triad components.

Since such operators are densely defined \cite{QSDV}, and sometimes
even bounded \cite{InvScale}, they cannot be identical to an inverse
of volume. In particular at small volume scales there are deviations
between the classical inverse and quantum behavior. While this
modified behavior is not determined uniquely, its characteristic
properties are robust.  Different versions arise because there are
many different ways to re-write inverse triad components, e.g.\ using
different representations for holonomies, which all give the same
classical expression but differ in quantum properties contained in
their spectra (see \cite{Spectra} for explicit examples). There are
thus quantization ambiguities, as always when one quantizes
expressions which are non-linear in basic quantities.

\subsection{Dynamics}

\begin{figure}[ph]
\centerline{\begin{picture}(50,15)(0,10)
\put(0,0){\vector(1,0){30}} \put(0,0){\vector(1,1){21.2}}
\put(30,0){\line(1,1){21.2}} \put(21.2,21.2){\line(1,0){30}}
\put(20,-5){\makebox(0,0){$\vec{s}_1$}}
\put(5,15){\makebox(0,0){$\vec{s}_2$}}
\put(25,10){\makebox(0,0){$\Delta$}}
\put(45,10){\makebox(0,0){$\alpha$}}
\end{picture}
\hspace{1cm}
\parbox{8cm}{\caption{Shape of loops used to express curvature components in terms
of holonomies. \label{Loop}}}}
\end{figure}
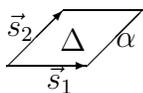

Inverse powers of densitized triad components are also necessary for
the Hamiltonian constraint (\ref{HamConstr}). The same type of
modifications thus results. Moreover, we have to express the curvature
components $F_{ab}^i$ in the first line of (\ref{HamConstr}) in terms
of holonomies before we can quantize. This can be done as usually in
gauge theories using $s_1^as_2^bF_{ab}^i\tau_i=
\Delta^{-1}(h_{\alpha}-1) +O(\Delta)$ for a holonomy around a closed
loop $\alpha$ of coordinate area $\Delta$ and with unit tangent vectors
$s_{1/2}$ as in Fig.~\ref{Loop}.

The second line of (\ref{HamConstr}), finally, requires us to quantize
the spin connection components which are complicated functionals of
the triad, and also contain their inverse. In general, it is easier to
quantize the combination $K_a^i=\gamma^{-1}(A_a^i-\Gamma_a^i)$
directly, which can again be written as a Poisson bracket \cite{QSDI}
\[
 K_a^i=\gamma^{-1}(A_a^i-\Gamma_a^i)
 \propto \left\{\!A_a^i,\!\left\{\int\md^3x F_{ab}^i
 \frac{E^a_jE^b_k}{\sqrt{|\det
 E|}},\int{\sqrt{|\det E|}}\mathrm{d}^3x\right\}\!\!\right\}
\]
while specific models may allow simpler direct quantizations of spin
connection components \cite{Closed,Spin}.

Also here, there are several quantization ambiguities and several
classes of operators. Any such construction results, understandably,
in an operator which is difficult to handle for explicit
calculations. It is thus helpful to look at symmetric models at the
quantum level \cite{SymmRed}, which display the characteristic
features. Such models, in turn, can then also provide feedback to the
full theory in order to guide or specify constructions there. In the
simplest models, isotropic ones \cite{IsoCosmo}, the basic variables
$c$ and $p$ are represented in a loop quantization on orthonormal
states $\langle c|\mu\rangle= e^{i\mu c/2}$ with real labels
$\mu\in{\mathbb R}$ \cite{Bohr}. The basic operators then act as
$\hat{p}|\mu\rangle= {\textstyle\frac{1}{6}}\gamma\ell_{\rm
P}^2\mu|\mu\rangle$ and $\widehat{e^{i\mu'c/2}}|\mu\rangle=
|\mu+\mu'\rangle$.

Following the construction of Hamiltonian constraint operators gives
\cite{IsoCosmo,Closed,Bohr}
\[
 \hat{H}|\mu\rangle =
 \frac{3}{16\pi G\gamma^3\ell_{\mathrm{P}}^2}
 (V_{\mu+1}-V_{\mu-1})
 (e^{-ik}|\mu+4\rangle
 -(2+k^2\gamma^2)|\mu\rangle+
 e^{ik}|\mu-4\rangle)
\]
where differences in volume eigenvalues $V_{\mu}=(\gamma\ell_{\rm
P}^2|\mu|/6)^{3/2}$ come from commutators, and the shifts in state
labels arise from operators $\widehat{e^{i\mu c/2}}$ used to quantize
$c^2$ in the classical constraint (\ref{Hamiso}). Expanding difference
operators in a Taylor series of differential operators then results in
higher order terms, i.e.\ higher powers of $c$ or $\dot{a}$ but no
higher spatial derivatives in homogeneous models, which become
important away from semiclassical regimes
\cite{SemiClass}. Similarly, the matter
Hamiltonian requires a quantization of $a^{-3}=|p|^{-3/2}$, while
$\hat{p}$ does not have a densely defined inverse. Re-writing as
before results in well-defined expressions which modify the classical
behavior on small scales (see \cite{Spectra} for explicit spectra).

\subsection{Sources for quantum corrections}

This outline of general constructions shows which types of corrections
we have to expect from loop quantum gravity if we use it to derive
effective equations from the expectation value of $\hat{H}$. There are
\begin{itemize}
\item modified coefficients due to the
 small-scale behavior of quantized inverse powers of triad
 components in matter Hamiltonians as well as the Hamiltonian
constraint and
\item higher order spatial
 derivatives which arise from replacing local curvature and connection
components by holonomies along extended loops; this also implies
spatial non-locality.
\end{itemize}
In addition to that, higher order time derivatives, although not
obvious from the loop constructions, result as per our general
discussion in the main text. There is no complete derivation yet, but
several studies for isolated specific corrections of different origin
have been performed. In general, one has to consider all of them, each
one to a certain order in perturbative treatments, since they have
different magnitudes depending on which regime is being
considered. This is similar to usual effective action pictures which
contain modified coefficients of different forms and also several
higher order and higher derivative corrections. Although the
derivation of a low energy effective action will be difficult since so
far no perturbative vacuum is known, the general procedure of
effective systems is applicable. The types of corrections expected
from loop quantum gravity are then completely analogous to what one
has for quantum field theories on a background. In particular, loop
quantum gravity does give rise to quantum corrections to the classical
Einstein--Hilbert action.

\section{Effective actions from string theory}

While it is difficult to define a perturbative vacuum or other state
corresponding to Minkowski space in loop quantum gravity, string
theory \cite{Polchinski} is originally defined perturbatively for string
worldsheets in a background spacetime. This formulation lends itself
directly to the computation of a low energy effective
action. Consistency of the theory then requires the introduction of
additional fields which also occur in the effective action coupled to
gravity. For gravity itself, higher curvature terms arise. This is
often not derived by the usual procedure to arrive at low energy
effective actions but by using conditions for the $\beta$-functions to
vanish in agreement with the preservation of conformal invariance of
the classical theory. This has the advantage of being applicable in
different background spacetime geometries.

Low energy effective actions are thus not just obtained around the
Minkowski vacuum but there are additional parameters to specify the
background geometry. This is in agreement with general expectations
formulated in Sec.~\ref{Impl}, but does not fully include parameters
for true quantum degrees of freedom such as spread. The derivation of
such an effective action would require more than perturbative
aspects. The available results are thus valid for low energy
properties of the theory but not necessarily for aspects of cosmology.

To compare with the results currently available for loop quantum
gravity, there is, as often, a very complementary picture: in one case
it is difficult to define a non-degenerate vacuum state while the
other is built around such a state and describing strong deviations
from this state is more involved. Nonetheless, both formulations agree
on the types of corrections that are expected in a general picture,
although calculations are not yet detailed enough for a quantitative
comparison. Coefficients of effective equations as well as covariance
properties can be quite different, which will allow interesting means
to compare features of both theories.

\subsection{Classical Strings}

In order to discuss the different approaches from string theory to the
effective field equations,  we first introduce the framework briefly
based on the standard references \cite{Polchinski} and
\cite{Green-Schwarz-Witten} in the field. String theory is based
on the premise that there must exist a theory of which general
relativity as well as the standard model of particle physics are low
energy limits. Although we will not discuss how the standard model
might arise, it is important to recall that the different standard
model fields of diverse spin, together with the spin two metric field
and many new yet unseen fields, are supposedly oscillation modes of a
string. The string action for a reparametrization invariant
2-dimensional Lorentzian ``worldsheet'' is given by the Nambu--Goto
action
\begin{equation}S_{NG}[X^\mu]=
-\frac{1}{2\pi\alpha'}\int_M\md\tau\md\sigma
\sqrt{-\det{h_{\alpha\beta}}}\,,
\end{equation}
where $h_{\alpha\beta}$ is the worldsheet metric induced by the
background metric $G_{\mu\nu}(X)$ of the ``target space,'' a higher
dimensional manifold with coordinates $X^\mu$ assumed to contain the
world as we see it.  A canonical quantization of the square root
Lagrangian is difficult, but can be avoided by introducing an
auxiliary worldsheet metric $\gamma_{\alpha\beta}$  and using
the equivalent Polyakov action
\begin{equation}
 S_P[X^\mu;\gamma_{\alpha\beta}]=-\frac{1}{4\pi\alpha'}\int_M\md
 \tau\md\sigma
 \sqrt{-\det{\gamma_{\delta\epsilon}}}\gamma^{\alpha\beta}G_{\mu\nu}\partial_\alpha
X^\mu\partial_\beta X^\nu \,.
\end{equation}

In a Minkowskian target space, $G_{\mu\nu}(X)=\eta_{\mu\nu}$, the
equations of motion are obtained by varying the Polyakov action with
respect to $X^\mu$, thus obtaining
\begin{equation}\partial_\alpha\sqrt{|\gamma|}\gamma^{\alpha\beta}
\partial_\beta X^\mu(\tau,\sigma)=0.
\end{equation} 
Solutions can be constructed once boundary conditions are specified,
for which there are periodic ones (closed string) or of Dirichlet,
Neumann or even mixed type where the positions of endpoints or their
derivatives are fixed. It is usual to fix all the gauge freedom in the
worldsheet metric and keep the conformal transformations as the only
degree of freedom left,
$\gamma_{\alpha\beta}(\tau,\sigma)=\omega^2(\tau,\sigma)\eta_{\alpha\beta}$.
In this conformal gauge, the equations are manifestly invariant under
changes of the conformal factor $\omega$.  In order to keep track of
the physical degrees of freedom of the theory, it is sometimes more
convenient to choose the so called light cone gauge where
$X^0(\tau,\sigma)+X^1(\tau,\sigma)=\tau$ and
$X^0-X^1=X^-$. If a proper coordinate transformation on the worldsheet
is performed, it is possible to write the Polyakov action in the
following shape
\begin{equation}S_{pp}=-\frac{1}{2\pi\alpha'}
\int\md\tau\int_0^{\ell}\md\sigma\left[
2\gamma_{\sigma\sigma}\partial_\tau X^-+
\gamma_{\sigma\sigma}\partial_\tau X^i\partial_\tau
X^i-\gamma_{\sigma\sigma}^{-1}\partial_\sigma X^i\partial_\sigma
X^i\right]
\end{equation} 
which for Neumann boundary conditions can be solved by
\begin{equation}
X^i(\tau,\sigma)=x^i+\frac{p^i}{p^+}\tau+i\sqrt{2\alpha'}
{\sum_{0\neq n=-\infty}^{\infty}}\frac{\alpha_n^i}{n}\cos\left(\frac{\pi
n\sigma}{\ell}\right)e^{-\frac{i}{\ell}\pi n c\tau}\,.
\end{equation} 
where $i\neq 0,1$ and $p^+$ is the average of the momentum canonically
conjugate to $X^-$. For other boundary conditions, one obtains
analogous situations with solutions decomposed in mode expansions
where the amplitude $\alpha_n^i$ of each mode is an independent degree
of freedom, with Poisson brackets
 $\left\{\alpha^i_m,\alpha_{-n}^j\right\}=-im\delta^{ij}\delta_{mn}$.

\subsection{Quantum Strings} 

The classical Lorentzian and conformal symmetries of string theory are
not necessarily preserved after quantization. This is only the case if
the target space is 26-dimensional (or 10-dimensional for
supersymmetric versions).  In particular conformal symmetry is the
cornerstone of string theory and its consistency as a physical theory
depends on it. It is also crucial as the source  of many mathematical
tools which may provide the calculational power that a good physical
theory should be equipped with.

Exact quantization can be performed only in special cases where the
string action becomes a free theory on the worldsheet.  This usually
requires a background metric of maximal symmetry among which Minkowski
and anti-de Sitter spacetimes have been most widely studied (see
\cite{Background} for general references on
strings in background fields). The modes $\alpha_n^i$ of the classical
string are then turned into operators, satisfying the Heisenberg
algebra $\left[\hat\alpha^i_m,\hat\alpha_{-n}^j\right]=
m\delta^{ij}\delta_{mn}$ where the reality condition of $X^\mu$
requires $(\hat\alpha_{-n}^j)^\dagger=\hat\alpha_{n}^j$. This operator
can be considered as a creation operator for the mode of rotation
labeled by $n$ corresponding to the spin of the created particle.

In addition, there are constraints which, when quantized,
obey a Virasoro algebra and annihilate physical states. 
The Hamiltonian then determines the mass spectrum, and is for the open
string given by
\[
H=\frac{1}{2p^+}p^ip^i+\frac{1}{2p^+\alpha'}
\left(\sum_{n>0}\alpha_{-n}^i\alpha_n^i+A\right)
\]
where regularization yields $A=\frac{2-D}{24}$ with $D$
the target space dimension. Therefore, the lowest mass state
can be constructed as $|0;k\rangle$ with squared mass
$m^2=\frac{2-D}{24\alpha'}$ and wave vector $k$. The vectorial mode
$\alpha_{-1}^i|0;k\rangle$ has $m^2=\frac{26-D}{24\alpha'}$, and
higher spin excitations have larger masses.

It has been argued that a theory including exclusively open string
modes is inconsistent, introducing closed string modes
\begin{equation}
X^i(\tau,\sigma)=x^i+\frac{p^i}{p^+}\tau+i\sqrt{\frac{\alpha'}{2}}
{\sum_{0\neq
n=-\infty}^{\infty}}\left(\frac{\bar\alpha_n^i}{n}e^{-\frac{2\pi i
n}{\ell}(\sigma+ c\tau)}+\frac{\tilde{\alpha}_n^i}{n}e^{\frac{2\pi i
n}{\ell}(\sigma- c\tau)}\right)\,.
\end{equation} 
Now, there is a constraint that removes any state with
different numbers of bar and tilde operators from the physical
spectrum. The lowest mass states are then for the
scalar degree of freedom $|\bar 0,\tilde 0;k\rangle$ with squared
mass $m^2=\frac{2-D}{6\alpha'}$ and the state
$\bar\alpha_{-1}^i\tilde\alpha_{-1}^j|\bar 0,\tilde 0;k\rangle$ with
$m^2=\frac{26-D}{6\alpha'}$ can be decomposed into the $SO(D-2)$
fundamental representations in the tensor product of two fundamental
ones.

Since self-consistency of the theory requires $D=26$, the fundamental
scalars are tachyons and the vector excitation, the antisymmetric
tensor, symmetric traceless tensor and trace part would all become
massless fields, usually regarded as a $U(1)$ connection, a
fundamental antisymmetric tensor ($B$-Field), the graviton field and a
dilaton scalar field.  Tachyonic states are absent in supersymmetric
versions, while the other excitations remain.  For our purposes of
illustrating effective corrections, we focus on the propagation of the
massless sector of the theory.

\subsection{Effective Field equations}

Since the squared mass spectrum is inversely proportional to the small
$\alpha'$, the only relevant particles in the low energy effective
dynamics of the string are the massless ones.  It is therefore of
interest to construct field equations which describe the propagation
of the string modes individually and, simultaneously, introduce
interactions between the effective fields at an effective level as
they follow from scattering amplitudes derived from the fundamental
string processes.  It turns out that self interactions (4-point
functions) of symmetric tensor modes correspond to those coming from
the linearized Einstein equations. Thus, very near to 26-dimensional
Minkowski spacetime the massless symmetric tensor excitations  of the
closed string behave as gravitational perturbations.

Field equations for each separate string mode can be
considered as the linearized expressions, corresponding to
two point functions. For higher order terms in
the expansion, it is then necessary to compute higher point
functions which is more difficult. 
In fact, relevant perturbative expansions have been carried out in the
$\alpha'$ expansion as well as in genus of the string worldsheet
topology, but most notorious is the inclusion of effects due
to the presence of highly massive states.  Their effects are included
by new interaction terms, similarly to a Fermi interaction.  Obviating
the need for other fields, an ansatz for an effective (low energy)
action can be considered as a sum of invariants constructed from
different powers of the Riemann tensor with free coefficients. Based
on properties of the different string theory models, these
coefficients in the low energy action can then be determined.

Parallel results have been performed  based on conformal field
theory techniques. These are very powerful and allow the study of the
unperturbed equations through the use of operator product expansions,
conformal invariance, the Virasoro algebra and vertex operators. In
order to obtain physical predictions
one deals with the $N$ particles scattering amplitude obtained
by an $S$-matrix. In these terms, it becomes very
useful to proceed by substituting the problem by a sum over all
possible configurations through the Feynman integral. In this
context it is possible to obtain an alternative viewpoint of the
pathology that arises in the scattering amplitudes which in the
lightcone gauge appears as an anomaly in the Lorentz invariance
unless $D=26$. In the present formalism the anomaly appears in the
conformal symmetry, which at the classical level  was realized as
$\gamma^{\alpha\beta}\frac{\delta }{\delta\gamma^{
\alpha\beta}}S_P[X^\mu]=\sqrt{\gamma}T^\alpha_{\ \alpha}=0$ but
holds no longer after quantization.

As a matter of fact, the scale invariance is evident due to the fact
that the Lagrangian contains only dimensionless coupling constants.
In spite of that, after quantization one obtains nonzero beta
functions which govern the flow of the parameters of the theory
according to the renormalization group equations.  The trace of the
energy momentum tensor then does not vanish after quantizing, and in
fact is proportional to the beta functions. They thus have to be set
to zero in order to make the results of physical predictions independent
of the fiducial metric $\gamma_{\alpha\beta}$.

Conformal field theory tools allow the computation of $\beta$
functions associated to each of the fields. For the gravitation mode,
the result is proportional to Einstein's equations. Therefore,
requiring that we are in the physical sector of string theory by
simply imposing that conformal invariance is preserved is equivalent
to imposing Einstein's equations.  These beta functions can in fact be
corrected to higher order in $\alpha'$ or in the genus expansion, thus
ideally leading to effective actions with higher
powers of the invariants of the theory such as $R^n$ although the
results depend on the type of string model.

\subsection{Free parameters}

From this brief review it is clear that techniques of string theory
are most suitable to low energy effective actions, but make it more
complicated to derive more general effective systems. Parameters
specifying states to expand around, as they occur in the general case,
will thus also be present for string theory, although current
technology does not allow one to include them. Some of those
parameters can be included by choosing different backgrounds, but this
does not include truly quantum variables. One can see directly how
these parameters are excluded by presently available calculations: As
we noted before, a general effective system will not be manifestly
invariant under all symmetries of the quantum theory because one has
to choose a state which may not be invariant. Similarly, setting
$\beta$-functions to zero implicitly selects special states used for
an effective system. This removes most of the freedom in choosing
states, similarly to picking explicitly the vacuum state for a low
energy effective action which preserves all symmetries manifestly.

\end{document}